\documentclass[review]{elsarticle}

\usepackage[colorlinks=true, allcolors=blue]{hyperref}
\usepackage[utf8]{inputenc}					%merkistö kuntoon eli ääkköset mukaan
\usepackage[T1]{fontenc}						%nätimmät ääkköset
\usepackage[]{units}

\journal{Nuclear Instruments and Methods}
\begin{document}

\begin{frontmatter}

\title{A new off-line ion source facility at IGISOL}

\author[a]{M. Vil\'en\corref{mycorrespondingauthor}}
\cortext[mycorrespondingauthor]{Corresponding author}
\ead{markus.k.vilen@student.jyu.fi}
\author[a]{L. Canete}
\author[b]{B. Cheal}
\author[c]{A. Giatzoglou}
\author[a]{R.	de Groote}
\author[a]{A.	de Roubin}
\author[a]{T.	Eronen}
\author[a]{S.	Geldhof}
\author[a]{A.	Jokinen}
\author[a]{A.	Kankainen}
\author[a]{I.D.	Moore}
\author[a]{D.A.	Nesterenko}
\author[a]{H.	Penttilä}
\author[a]{I.	Pohjalainen}
\author[a]{M.	Reponen}
\author[a]{S.	Rinta-Antila}

\address[a]{University of Jyv{\"a}skyl{\"a}, P.O. Box 35, FI-40014 University of Jyv{\"a}skyl{\"a}, Finland}
\address[b]{Department of Physics, University of Liverpool, Liverpool L69 7ZE, United Kingdom}
\address[c]{Department of Physics and Astronomy, University College London, Gower Street, London WC1E 6BT, United Kingdom}

\begin{abstract}
An off-line ion source station has been commissioned at the IGISOL (Ion Guide Isotope Separator On-Line) facility. It offers the infrastructure needed to produce stable ion beams from three off-line ion sources in parallel with the radioactive ion beams produced from the IGISOL target chamber. This has resulted in improved feasibility for new experiments by offering reference ions for Penning-trap mass measurements, laser spectroscopy and atom trap experiments. 
\end{abstract}


\end{frontmatter}

%\linenumbers

\section{A new ion source facility at IGISOL-4}
\begin{figure}
\centering
\includegraphics[trim = 0mm 0mm 0mm 0mm, clip, width=0.9\textwidth]{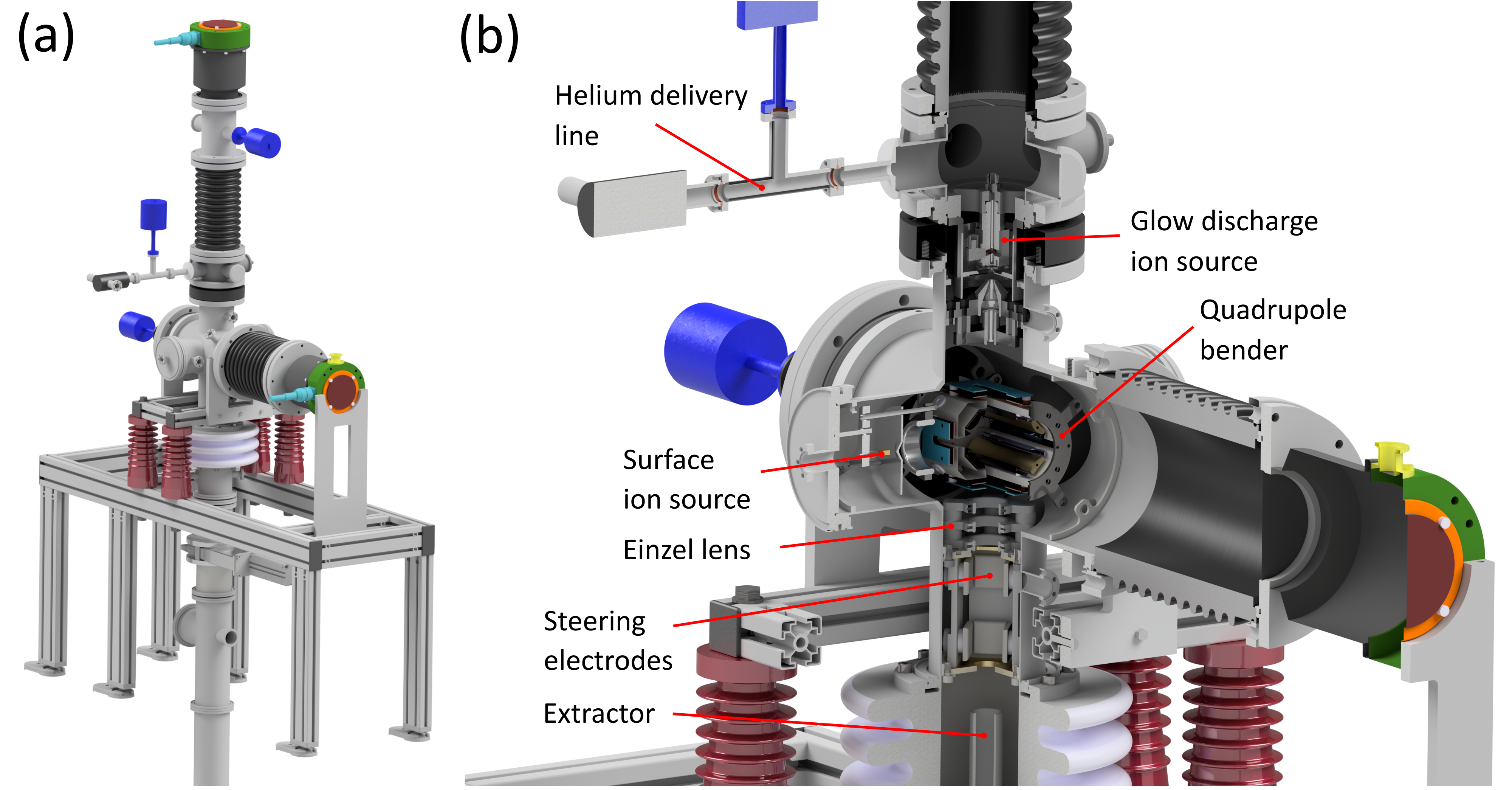}
\caption{The new off-line ion source station (a) and its internal structure (b).}
\label{is_station}
\end{figure}

In this contribution, we present the latest addition to the IGISOL-4 (Ion Guide Isotope Separator On-Line) facility \cite{moore2013}, a new off-line ion source station, see figure \ref{is_station}. The new ion source infrastructure consists of an ion source setup located on the second floor of the experimental hall and a beamline connecting the new system to the main IGISOL mass separator. The ion source station is designed to accommodate three ion sources simultaneously, two sources in the horizontal branches of the vacuum system and one vertically mounted. 

The new station houses electrostatic ion optics that can accept a beam from each of the three directions. Both ion sources presented in figure \ref{is_station} utilize a skimmer electrode to form an ion beam which is injected into the quadrupole bender. The beam is focused with an Einzel lens, situated immediately after the bender, and its alignment is adjusted using a set of steering electrodes. The ions have $\unit[800q]{eV}$ of energy, $q$ being the charge state of the ions, before entering the extractor electrode which further accelerates them to $\unit[30q]{keV}$.

Subsequently, ions are transported to the lower floor of the IGISOL facility, deflected by $90^\circ$ into the horizontal beamline and separated based on their $m/q$ ratio using a dipole magnet. The bending optics have been designed to accept beam from either the off-line ion source station or the IGISOL target chamber. Switching between these two sources is achieved by alternating a steering electrode voltage within the bender.

In its present configuration, the ion source station houses two types of ion source, a surface ion source based on Ohmic heating and a glow discharge ion source. The former ionizes a variety of alkali earth and alkali metal elements, whereas the latter source is a more flexible device that ionizes its cathode material via electric discharge through helium buffer gas. The ion source station is equipped with a cryogenic buffer gas purification system in order to provide beams with higher concentrations of the cathode material. Commissioning of a third ion source, a laser ablation ion source, is being planned.

\section{Commissioning of the ion source station}

After the first commissioning runs with $^{63}\mathrm{Cu}$ and $^{133}\mathrm{Cs}$, the new ion source station has been employed in many experiments at IGISOL. The station was used to provide ion samples in parallel to the IGISOL system. This enabled the laser spectroscopy work in \cite{vormawah2018} by making $^{89}\mathrm{Y}^{2+}$ ions available where they would not have been previously and made possible the Penning trap mass measurement campaign, see \cite{nesterenko2019}, conducted fully off-line using both beamlines. Additionally, the ion source station has been used to provide well-known and identified reference ions for the on-line Penning trap mass measurements in \cite{vilen2019} and to provide $^{133}\mathrm{Cs}^{+}$ for demonstrating the operation of a new laser cooling and trapping facility for the production of ultra-cold atomic samples of caesium \cite{GIATZOGLOU2018}. The new station has been proven to be a useful tool in preparing for on-line experiments through allowing initial tuning of both Penning trap and collinear laser systems without disrupting preparation of the IGISOL front end.

Operating parameters of the system were mostly the same during these experiments, with the exception of ion source settings. Typical values for the glow discharge ion source were $\unit[5-10]{mbar}$ helium pressure and $\unit[800-1400]{V}$ source voltage. Current used to heat the surface ion source was adjusted based on the amount of beam needed, typically being $\unit[1.5-1.8]{A}$.

\section{Conclusions}

The new ion source infrastructure has been used in several experiments demonstrating additional flexibility in the operating modes of the IGISOL facility, enabling novel experiments and the possibility to conduct Penning trap mass measurement campaigns in a fully off-line operating mode. The off-line ion source station, with its two commissioned ion sources, provides a large variety of mass separated ion beams to experiments. Commissioning of a third ion source based on laser ablation would not only provide an increased number of ion species for experiments, but also a possibility to explore carbon cluster formation which would be of interest as reference ions for mass measurements.

%\bibliography{EMIS_proceedings_2018_bibliography}

\end{document}